\documentstyle[prl,aps,epsfig]{revtex}

\newcommand{\commento}[1]{}

\newcommand{\be}{\begin{equation}}
\newcommand{\ee}{\end{equation}}
\newcommand{\beq}{\begin{eqnarray}}
\newcommand{\eeq}{\end{eqnarray}}

\newcommand{\cu}[0]{\mathrm{CuO}_{4} }

\begin{document}
   
\def\gC{\mbox{\boldmath $C$}}
\def\gZ{\mbox{\boldmath $Z$}}
\def\gR{\mbox{\boldmath $R$}}
\def\gN{\mbox{\boldmath $N$}}
\def\ua{\uparrow}
\def\da{\downarrow}
\def\a{\alpha}
\def\b{\beta}
\def\g{\gamma}
\def\G{\Gamma}
\def\d{\delta}
\def\D{\Delta}
\def\e{\epsilon}
\def\ve{\varepsilon}
\def\z{\zeta}
\def\h{\eta}
\def\th{\theta}
\def\k{\kappa}
\def\l{\lambda}
\def\L{\Lambda}
\def\m{\mu}
\def\n{\nu}
\def\x{\xi}
\def\X{\Xi}
\def\p{\pi}
\def\P{\Pi}
\def\r{\rho}
\def\s{\sigma}
\def\S{\Sigma}
\def\t{\tau}
\def\f{\phi}
\def\vf{\varphi}
\def\F{\Phi}
\def\c{\chi}
\def\w{\omega}
\def\W{\Omega}
\def\Q{\Psi}
\def\q{\psi}
\def\de{\partial}
\def\inf{\infty}
\def\ra{\rightarrow}
\def\bra{\langle}
\def\ket{\rangle}

\draft

\twocolumn[\hsize\textwidth\columnwidth\hsize\csname
@twocolumnfalse\endcsname

\widetext

\title{Pairing in Cu-O Models: Clues of Joint  Electron-Phonon  and Electron-Electron 
Interactions}

\author{Enrico Perfetto and Michele Cini}
\address{Istituto Nazionale di Fisica Nucleare, Dipartimento di
Fisica,\\
Universita' di Roma Tor Vergata, Via della Ricerca Scientifica, 1-00133\\
Roma, Italy}

\maketitle

\begin{abstract}

  We discuss a  many-electron Hamiltonian with Hubbard-like 
    repulsive interaction and linear coupling to the   phonon branches, 
    having the Cu-O plane of the superconducting cuprates as a paradigm.  
    A canonical transformation  extracts an effective two-body problem from 
    the many-body theory. As a prototype system we study the $\cu$ cluster, 
    which yields 
    electronic  pairing in the Hubbard model; moreover,  a standard treatment of 
    the Jahn-Teller 
   effect  predicts distortions that destroy electronic 
    pairing. Remarkably, calculations that keep all the electronic 
    spectrum into account show that vibrations are likely to be synergic with electronic 
    pairing, if the coupling to half-breathing modes predominates, as 
    experiments suggest.

\end{abstract}

\bigskip\bigskip\bigskip
]

\narrowtext

{\small

The role of electron-phonon (EP) interactions in determining the
superconducting correlations in the Cu-O planes of cuprates  is a very 
controversial issue. Possibly, the pairing mechanism 
 has a predominantly electronic origin
\cite{dagotto}, but many high-$T_{C}$ compounds exhibit a quite 
noticeable doping dependent isotope effect\cite{crowford}, suggesting that EP 
interactions could be important and should be included in the theory.
In particular there is experimental evidence\cite{mcqueeney} that the
half breathing Cu-O bond stretching mode at $k=(\pi,0),(0,\pi)$ is significantly coupled 
with the doped holes in the superconducting regime and its 
contribution may be relevant for the $d_{x^{2}-y^{2}}$ 
pairing\cite{lanzara}\cite{gunnarsson}\cite{ishihara}. 

The simplest way to include both strong electronic correlations 
and  EP interactions is the Hubbard-Holstein 
model, where electrons are coupled to a local Einstein phonon.
Much is known\cite{pao}\cite{alexandrov}
about the possibility of a  superconducting phase in this model;  
 further evidence for pairing can be extracted by exact 
 numerical diagonalization of 
small cluster Hamiltonians, calculating the effective pairing interaction 
defined as ${\tilde \Delta} (N+2)=E(N+2)+E(N)-2E(N+1)$,
where $N$ is the number of fermions in the system.
A negative ${\tilde \Delta}$ is interpreted as an effective attraction and 
suggests the presence af a bound pair in the ground state.
Petrov and Egami\cite{egami} found ${\tilde \Delta} <0$ in a 
doped 8-site Hubbard-Holstein ring and showed that the effect disappears once 
phonons are turned off. 
Conversely Mazumdar and coworkers\cite{mazumdar} suggested that 
pairing in Hubbard clusters reported by some authors\cite{white}
was of doubtful physical  interpretation due to the 
neglect  of the lattice degrees of freedom and the  Jahn-Teller (JT) effect. 
Consider even $N$, such that the $N+2$-particle cluster hosts a singlet 
pair and  the $N+1$-particle ground state is degenerate. 
They argued that JT distorsion might cause a larger  energy gain of the 
system with $N+1$ particles, thus reversing the sign of ${\tilde \Delta}$.
This led to the conjecture that any ${\tilde \D}<0$ due to 
an electronic mechanism is just a finite size  effect, which vanishes for large 
systems like the JT effect does. 

Here we further investigate this issue by addressing the question if the
recently proposed $W=0$ pairing avaliable in the Hubbard 
model\cite{PRB1997}\cite{PRB2002}\cite{PRB2003} survives when the
lattice degrees of freedom are switched on.
In  previous works it was shown that a class 
of $C_{4v}$-symmetric clusters exhibit the ${\tilde \D}<0$ property, 
due to the binding of the so called $W=0$ pairs. This depends crucially on the symmetry and on 
the availability of degenerate states.  Fermions in strongly correlated systems can avoid the on-site 
repulsion and possibly form singlet bound states if there is enough symmetry; 
the fillings and the symmetry channels where the $W=0$ pairing can occur are determined
in full generality by the $W=0$ theorem\cite{IJMPB2000}; these 
symmetries achieve  
the same result as high  angular momentum and parallel spins in the 
Kohn-Luttinger\cite{kohn} continuum approach. 

When lattice effects are introduced in this scenario,  several questions 
arise. In the conventional  mechanism, phonons overscreen 
the electron repulsion; what happens if electronic screening already 
leads to pairing? It is not obvious that  the phonons will reinforce the attraction while 
preserving the symmetry.  More generally, some vibrations could be pairing and 
others pair-breaking.   To address these problems we use an 
extension of the Hubbard model in which bond stretchings dictate the 
couplings to the normal modes of the $C_{4v}$-symmetric configuration, 
generating a long-range (Fr\"ohlich) EP interaction. This is 
physically more detailed than the  Hubbard-Holstein model, and does not 
restrict to on-site EP coupligs that would be impaired by a
strong Hubbard repulsion. A further  important drawback of the 
 Holstein EP interaction in this context is an unphysically large polaron 
 (and bipolaron) mass and low  polaron mobility. This problem is avoided using Fr\"ohlich-like 
 phonons  for modeling the Cu-O  planes. It was pointed out 
 recently\cite{bonca}\cite{alexandrov2}\cite{alexandrov3} 
that even  in the strong
EP coupling regime the  a long-range EP interaction removes the problem of  polaron 
self-trapping.   

We start from the  Hubbard model with on-site interaction $U$ and 
expand the hopping integrals $t_{i,j}({\bf r}_{i},{\bf r}_{j})$ in 
powers of the 
displacements ${\bf \r}_{i}$ around a $C_{4v}$-symmetric equilibrium configuration  
\begin{eqnarray}
t_{i,j}({\bf r}_{i},{\bf r}_{j}) \simeq t^{0}_{i,j}({\bf r}_{i},{\bf r}_{j})
+ \sum_{\a}\left[ \frac{\partial t_{ij}({\bf r}_{i},{\bf r}_{j}) }
{ \partial r^{\a}_{i}} \right] _{0}   \r^{\a}_{i}+\nonumber\\
+ \sum_{\a} \left[ \frac{\partial t_{ij}({\bf r}_{i},{\bf r}_{j}) }
{ \partial r^{\a}_{j}} \right] _{0}  \r^{\a}_{j}  
\; ,
\label{varhop}
\end{eqnarray}
where  $\a =x,y$. 
Below, we  write down the  $\r^{\a}_{i}$  in terms of the normal modes  
$q_{\eta \, \nu}$:
$
\r^{\a}_{i}=\sum_{\eta \, \nu} S^{\a}_{\eta \, \nu}(i) \; q_{\eta \, \nu} 
$, where $\eta$ is the label of   an irreducible representation ({\em irrep})  of the 
symmetry group of the undistorted system and $\n$ is a phonon branch.

Thus,  treating  the  Cu  atoms as fixed, for 
simplicity, one can justify an electron-lattice Hamiltonian:
\begin{equation}
H_{el-latt} =  H_{0} + V_{\rm tot} \, . 
\label{htot}
\end{equation}
Here $ H_{0} = H_{0}^{n}+H_{0}^{e}$ is given by
\begin{equation}
  H_{0} =\sum_{\eta} \hbar \omega_{\eta,\n} 
  b^{\dagger}_{\eta,\n} 
  b_{\eta,\n}+
  \sum_{i,j\s} t^{0}_{i,j}({\bf r}_{i},{\bf r}_{j})( c^{\dag}_{i 
  \s}c_{j \s}+h.c),
  \end{equation}
where $\omega_{\eta,\n}$ are normal mode 
  frequencies.
Moreover, let $M$ denote  the O mass, $\xi_{\eta,\nu}=\lambda_{\eta  \, 
  \nu}\sqrt{\frac{\hbar}{2M\omega_{\eta,\n}}}$, with $\lambda_{\eta  \, \nu}$ 
  numbers of order unity 
that modulate the EP coupling strength.
Then, $ V_{\rm tot} = V+W$   reads
  \begin{eqnarray}
  V_{\rm tot}=\sum_{\eta,\n} 
  \xi_{\eta,\nu} ( 
b^{\dagger}_{\eta,\n}+ b_{\eta,\n}) H_{\eta, \n}+
U\sum_{i}n_{i\ua} n_{i\da},
\label{hel}
\end{eqnarray}

the  $H_{\eta,\n}$ operators  are given by
\begin{eqnarray}
 H_{\eta,\n}=\sum_{i,j} \sum_{\a, \sigma} \left\{ S^{\a}_{\eta \, \nu}(i)
 \left[ \frac{\partial t_{ij}({\bf r}_{i},{\bf r}_{j}) }
{ \partial r^{\a}_{i}} \right] _{0}  \right. \nonumber\\ \left.+ S^{\a}_{\eta \, \nu}(j)
\left[ \frac{\partial t_{ij}({\bf r}_{i},{\bf r}_{j}) }
{ \partial r^{\a}_{j}} \right] _{0}    \right\} ( c^{\dag}_{i, 
\s}c_{j ,\s}+h.c.) \, .
\label{heta}
\end{eqnarray}

Consider the Hubbard model   $H_{H}= H_{0}^{e}+W$ involving the purely electronic terms of $H_{el-latt}$. 
We recall that $W=0$ pairs are defined as 
two-body singlet eigenstates of the $H_{H}$ free of double occupancy; in the many-body  
problem they interact indirectly and can get bound\cite{PRB1997}\cite{EPJB1999}. 
Here we wish to derive  an 
effective interaction between the particles in the pair suitable for  $H_{el-latt}$,
by   generalizing 
the canonical transformation approach of Ref.\cite{EPJB1999}.

Denote the phonon vacuum by  $|0\rangle \rangle$ and  the $N$-particle 
non-interacting Fermi sphere by $|\Phi_{0}(N) \rangle $; if we add a 
$W=0$ pair to
$|\Phi_{0}(N) \rangle \otimes |0\rangle \rangle$, the two extra particles, 
by definition, cannot 
interact directly (in first-order). 
Hence their effective interaction
comes out from virtual electron-hole (e-h) excitation and/or phonon exchange 
and in principle can be attractive.  
To expand 
 the interacting $(N+2)$-fermions 
ground state $|\Psi_{0}(N+2) \rangle$,
we build a complete set $\cal{S}$ of configurations in the subspace
with vanishing  $z$ spin  component, considering the  vacuum 
state 
  $|\F_{0}(N)\rangle \otimes |0\rangle \rangle $ 
and  the   set   of excitations over it. 

We start by creating  $W=0$  pairs of  fermions over
$|\F_{0}(N)\rangle \otimes |0\rangle \rangle$; we 
denote with $|m\ket \otimes | 0 \rangle \rangle$ these states.
At weak coupling,   we  may  truncate the Hilbert space to the 
simplest excitations, i.e., to states involving 1 e-h pair or 1 
phonon created over  the $|m\ket \otimes | 0 \rangle \rangle$
states. 
We define the $|m\ket \otimes | q \rangle \rangle$ states,
obtained by creating a phonon denoted by $q=(\eta,\nu)$ over
the $|m\ket \otimes | 0 \rangle \rangle$ states. 
Finally we introduce
the $ |\alpha \rangle  \otimes | 0 \rangle \rangle$ states,
obtained from the $|m\ket \otimes | 0 \rangle \rangle$ states by creating 
1 electron-hole (e-h) pair.   We now expand the interacting ground state in the truncated 
Hilbert  space: 
\begin{eqnarray}
|\Psi _{0}(N+2)\ket={\sum_{m}}a_{m}|m\ket \otimes |0 \rangle \rangle 
+\nonumber\\ + {\sum_{m,q}}a_{m,q}|m\ket \otimes |q \rangle \rangle 
+{\sum_{\alpha }}
a_{\alpha }|\alpha \ket \otimes |0 \rangle \rangle  
\label{lungo}
\end{eqnarray}
and set up the Schr\"{o}dinger equation with energy eigenvalue $E$. 
A canonical transformation\cite{pcfuturo} decouples the 
higher amplitudes $a_{m,q}$ and $a_{\a}$, and the $a_{m}$  become 
expansion  coefficients over the $W=0$ pairs of the wave function of the dressed pair $|\varphi \rangle$.
This obeys the Cooper-like equation   $H_{\rm pair}|\varphi \rangle  = E |\varphi \rangle  $
with an effective two-body Hamiltonian 
$H_{\rm pair}$ :
\begin{equation}
H_{\rm pair} \equiv \left(H_{0}+W+S[E]\right) \,
\label{sceq}
\end{equation}
where $S$ is the $E-$dependent effective scattering operator; the 
pairing problem must be solved self-consistently. 
It turns out that the effective  scattering operator  is
\begin{eqnarray}
S[E]_{m,m'}= \sum_{\a}
\frac{W_{m,\a}W_{\a,m'}}{E'_{\a}-E} +
\sum_{m'',q}\frac{V^{q}_{m,m''}V^{q}_{m'',m'}}
{E'_{m''}+\omega ' _{q}-E} \, .
\label{coopeff}
\end{eqnarray}
Here the primed quantites in the denominators are the  
eigenenergies of the $|m\ket \otimes |0 \rangle \rangle $, $|m\ket \otimes 
|q \rangle \rangle $,
$| \a \ket \otimes |0 \rangle \rangle $ states; they get renormalized 
in the decoupling procedure.
The pairing criterion involves the properly renormalized  Fermi energy $\varepsilon_{F}^{R}$ (see 
ref.\cite{PRB2002}); if the lowest energy eigenvalue $E$ is such that 
$E=2\varepsilon_{F}^{R}+\D$ with negative 
$\D$, the dressed $W=0$ pair gets bound in the many-body 
interacting problem and the system undergoes a  Cooper instability.  

As an illustrative application of the above pairing scheme, in this preliminary 
work  we focus  on CuO$_{4}$, the smallest cluster yielding $W=0$ pairing 
in the Hubbard model. This requires 4 holes, (total number, not 
referred to half filling); such a doping is somewhat unrealistic, but  larger 
$C_{4v}$-symmetric clusters and the full CuO$_{2}$ plane also show $W=0$ pairing in 
the doping regime relevant for cuprates\cite{PRB1997}\cite{EPJB1999}. 
Remarkably, in the pure Hubbard model, one can verify that $\D=\tilde{\D}(4)$ at least at 
weak coupling\cite{PRB1997}, which demonstrates that $\tilde{\D}$ has 
the physical meaning of an effective interaction. 
CuO$_{4}$  represents a  good test of the interplay 
between electronic and phononic pairing 
mechanisms since we can compare exact diagonalization 
results with the analytic approximations of the canonical 
transformation. CuO$_{4}$ allows   only the coupling to phonons at the centre 
or at  the edge of 
the Brillouin Zone;  however, phonons near the edge are precisely those  most 
involved\cite{mcqueeney}\cite{lanzara}.  
\begin{figure}[H]
\begin{center}
       \epsfig{figure=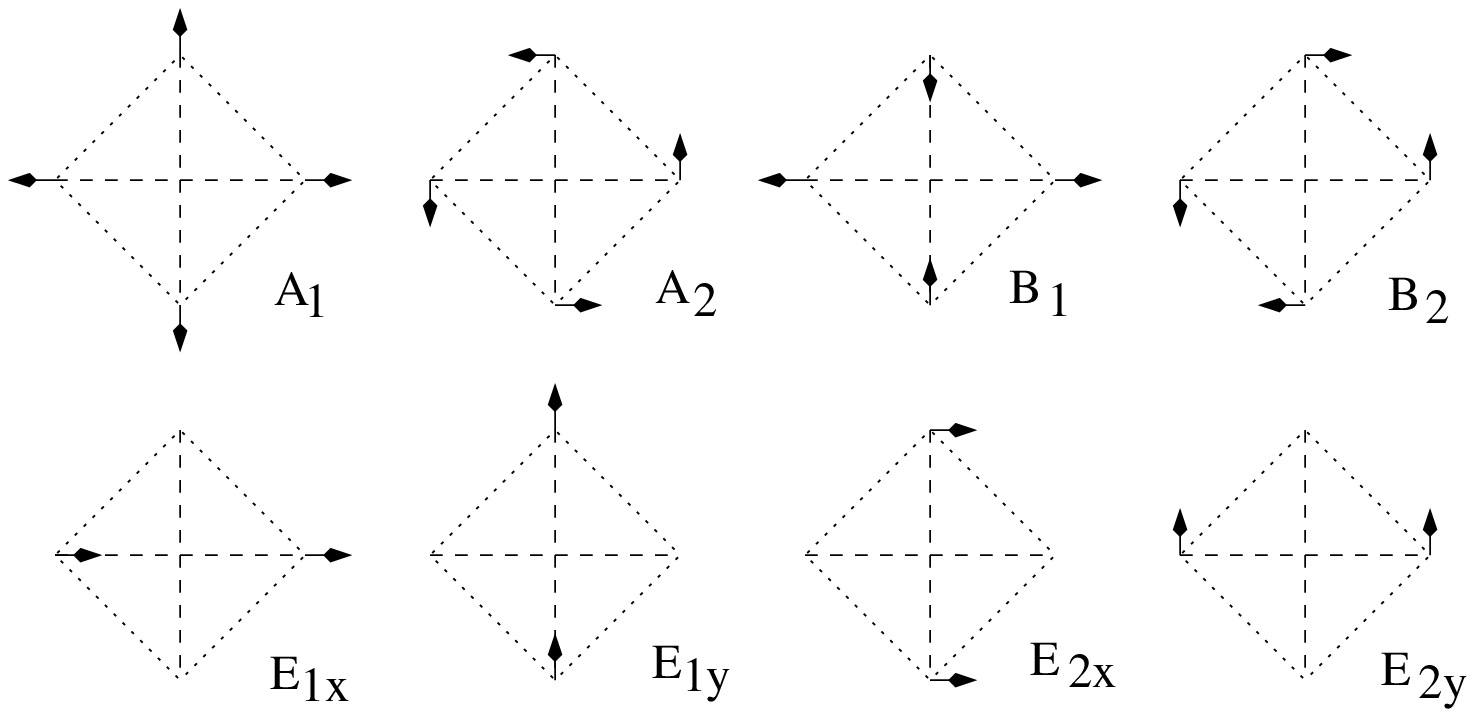,width=8cm}
       \caption{\footnotesize{
       Pictorial representation of the ionic displacements in the eight normal modes of the CuO${4}$ 
       cluster, labelled according to the {\em irreps} of the $C_{4v}$ Group. }}
 \label{cuo4}
\end{center} 
\end{figure}
Even in this small system the virtually 
exact diagonalizations are already hard and the next 
$C_{4v}$-symmetric example, the 
Cu$_{5}$O$_{4}$ cluster\cite{EPJB2000},
is  much more demanding for the number of vibrations and the size 
of the electronic Hilbert space.  The pure Hubbard CuO$_{4}$ cluster with O-O 
hopping    $t_{ox}=0$ yields\cite{PRB1997} ${\tilde \Delta} (4) <0$, 
due to a couple of degenerate   $W=0$ bound pairs,  in the $A_{1}$   
and  $B_{2}$ {\em irreps} of the $C_{4v}$ group; therefore in Eq.(\ref{coopeff}) we
set the $m=m'$ 
labels accordingly.
Starting with the $C_{4v}$-symmetric arrangement, 
any   displacement of the Oxygens in the plane can be analised in 
{\em irreps}, 
$A_{1},A_{2}, B_{1}, B_{2}, E_{1},E_{2}$, see Fig.\ref{cuo4}.
The EP coupling is expressed through two parameters $\partial t$ and  
$\partial t_{ox}$, defined as the derivatives of $t$ and $t_{ox}$ with 
respect to the Cu-O and O-O bond lengths.
The CuO$_{4}$ cluster is also interesting as a test of the 
conjecture of Ref. \cite{mazumdar}; we checked\cite{pcfuturo} that   
a standard JT 
calculation in which the degenerate ground state wave functions 
interact with the vibrations  indeed predicts distortions that already 
at moderate EP coupling  
destroy   $W=0$ pairing; here we wish to go beyond this approximation.

According to Eq.(\ref{coopeff}), the phonon-mediated 
interaction,  in the $A_{1}$ channel 
is:
\begin{eqnarray}
\sum_{m'',q}\frac{V^{q}_{A_{1},m''}V^{q}_{m'',A_{1}}}
{E_{m''}+\omega ' _{q}-E} 
=  -\frac{4}{3}\partial t_{ox}^{2}
\left( \frac{\lambda_{B_{2}}^{2}}{2\varepsilon_{A_{1}}+\omega_{B_{2}}-E} +           
\right. \nonumber\\ \left. 
\frac{2 \lambda_{A_{1}}^{2}}{2\varepsilon_{A_{1}}+\omega_{A_{1}}-E}
-\frac{\lambda_{E_{1}}^{2}}{2\varepsilon_{A_{1}}
+\omega_{E_{1}}-E}-\frac{\lambda_{E_{2}}^{2}}{2\varepsilon_{A_{1}}+\omega_{E_{2}}-E}\right). 
 \label{phona1}\end{eqnarray}
In the  $B_{2}$ channels we find:
\begin{equation}
\sum_{m'',q}\frac{V^{q}_{B_{2},m''}V^{q}_{m'',B_{2}}}
{E'_{m''}+\omega ' _{q}-E} 
=-4\partial t_{ox}^{2}
\frac{\lambda_{B_{2}}^{2}}{2\varepsilon_{A_{1}}+\omega_{B_{2}}-E} \;.
\label{phonb2}
\end{equation}

Thus, the two $W=0$ pairs  behave differently: the $B_{2}$
binding energy is enhanced by phonons, while in the $A_{1}$ sector
the overall sign depends on $\lambda_{\eta}$ and $\w_{\eta}$. It turns out that $A_{1}$ and $B_{2}$ modes 
are synergic to the $W=0$ pairing, while both longitudinal and 
transverse $E$ modes are 
pair-breaking.  The half-breathing  modes that are deemed most 
important\cite{lanzara}\cite{gunnarsson} are $A_{1}\pm B_{1}$ 
combinations, but $B_{1}$ is scarcely relevant.
For the sake of argument, in the explicit calculations we took all the normal modes 
with the same  energy 
$\varepsilon_{0}=\hbar \omega_{0} = 10^{-1} \rm{eV}$ and 
$\lambda_{\eta} = 1$. 
This  sets the  length scale of lattice effects
$\xi_{0}=\sqrt{\frac{\hbar}{2M \omega_{0}}}\simeq 10^{-1} \stackrel{\circ}{\rm{A}} 
$
where we used $M=2.7 \times 10^{-26}$Kg for Oxygens.
\begin{figure}[H]
\begin{center}
	\epsfig{figure=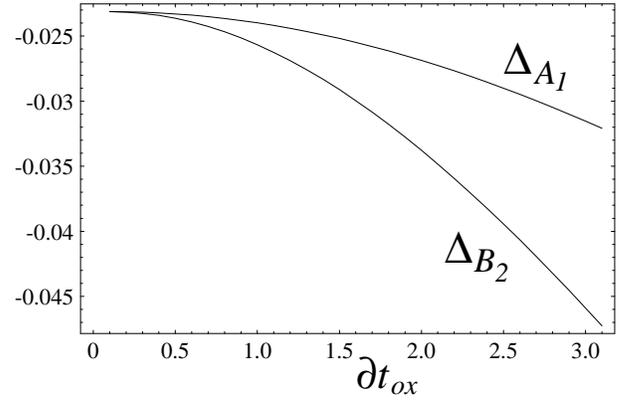,width=8cm}
	\caption{ \footnotesize{ Analytical results of the canonical transformation: 
	pair binding energy in the $A_{1}$ and $B_{2}$ sectors as a 
	function of $\partial t_{ox} $.
	Here we used $\lambda_{\eta}=1$ for every mode, $t=1$eV, $t_{ox}=0$, $U=1$eV;
	$\partial t_{ox} $ is in  units of  
	$\varepsilon_{0}/\xi_{0}=1$eV$\times 
	\stackrel{\circ}{\rm{A}}^{-1} $, $\D$ is in eV.}}
  \label{bind}
\end{center} 
\end{figure}
With this choice, pairing is enhanced in the $A_{1}$ sector as 
well, albeit less than in $B_{2}$, see Fig.\ref{bind}, where the 
pairing energy without phonons is $\D \simeq -20$meV for both the $W=0$ 
pairs.  In other 
terms, the vibrations split the degeneracy of the electronic ground 
state, effectively lowering the symmetry like a nonvanishing  
$t_{ox}$. Here again we stress that $\D=\tilde{\D}(4)$ for both 
symmetries, generalizing the Hubbard model result.
To go beyond the weak coupling case,  we diagonalized  $H_{el-latt}$  numerically   
in the truncated Hilbert space with up to 5 modes 
at a time having  vibrational  quantum numbers $n_{v} \leq 3$.
In Figure  $\ref{due}$a we included all vibrations except 
$E_{1}$;  with increasing the EP coupling, the pair 
binding energy steadily increases, in agreement with the trend in 
Fig.\ref{bind}; the unbinding mode $E_{2}$ is dominated by the 
pairing ones. Note that the binding energy increases with $\partial 
t_{ox}$ as predicted by the weak coupling theory, but also with  $|\partial 
t|$, suggesting that at strong coupling the Cu-O bond stretching grows 
in importance. 
\begin{figure}
\begin{center}
	\epsfig{figure=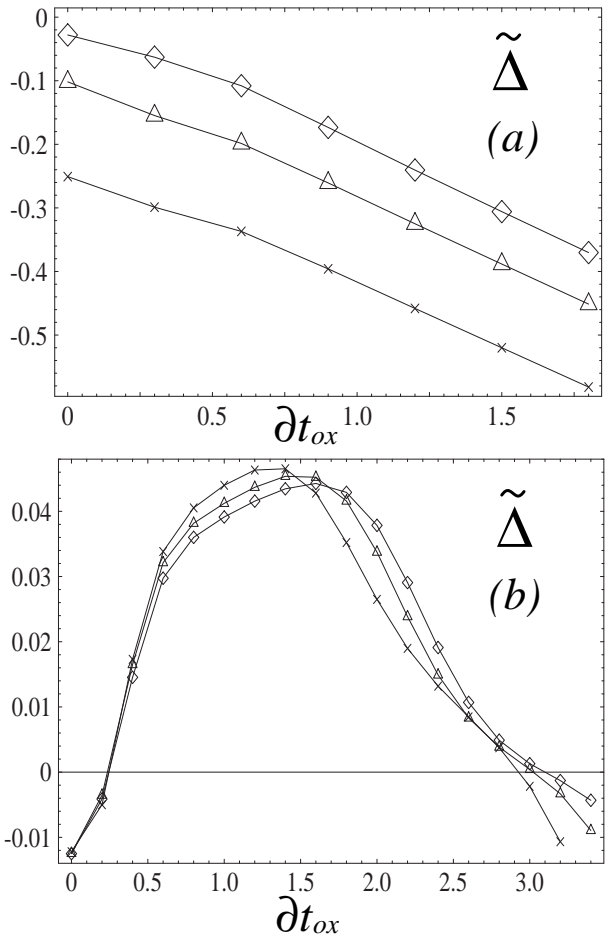,width=8cm}
	\caption{\footnotesize{${\tilde \D}(4)$ as a function of $\partial t_{ox}$ for different
	values of $\partial t$.
	  $\lambda_{\eta}=1 $ all the 
	vibrations, except:  $\lambda_{\eta}=0 $ for $\eta= E_{1}$ (a);
  $\lambda_{\eta}=0 $ for $\eta= E_{2}$ (b). 
	$n_{v}$=3 for each active mode;
	$\partial t=-0.2$ (diamonds);
	$\partial t=-0.5$ (triangles);
	$\partial t=-1$ (crosses).
	Here we used $t=1$eV, $t_{ox}=0$, $U=1$eV, $\partial t_{ox} $ and
	$\partial t$  are in  units of  
	$\varepsilon_{0}/\xi_{0}=1$eV$\times 
	\stackrel{\circ}{\rm{A}}^{-1} $,  $\tilde{\D}$ is in eV. }}
  \label{due}
\end{center} 
\end{figure}

In Figure  \ref{due}b we included all vibrations except 
$E_{2}$. The  $E_{1}$  vibrations are strongly pair 
breaking, and  $\tilde {\D}(4)$ is much smaller in magnitude than in 
Figure  
\ref{due}a; it becomes positive at intermediate coupling.  
It is likely that the couplings to the  pair-breaking $E$ modes are 
somewhat  underestimated by the choice of $\lambda$ parameters 
in Fig.\ref{due}a  and overestimated in Fig.\ref{due}b.
Interestingly,  however, at stronger coupling the pairing modes prevail 
again driving the system to pairing
(bipolaron formation). This resembles the trend found in Ref\cite{egami}, 
although there a Hubbard-Holstein framework is used and since the 
system is one-dimensional 
the electronic pairing does not occur.

These results yield several indications. First, the catastrophic distortions 
predicted by the  JT
approximation are not borne out by  more complete 
approaches involving  a broader spectrum of electronic states. This 
interesting result arises since the electronic ground state belongs to a 
narrow multiplet; the standard treatment of the JT effect singles out the lowest level, which is justified only at weak 
EP coupling.  Pairing prevails also at strong  coupling 
in part of the parameter space, in the symmetry channels where $W=0$ 
pairs occur. The correct trend is predicted by the canonical 
transformation approach, which also explains the pairing or 
pair-breaking character of the modes. In particular it is found that 
the half-breathing modes give a  synergic contribution to the purely 
electronic pairing; since   they are believed to be   mainly involved in 
optimally doped cuprates, our findings suggest  a joint 
mechanism, with the Hubbard model that captures a crucial part of the physics.
This scenario was also drawn in the context of an extended t-J model, where  
the half-breathing mode was found to enhance  electronic pairing\cite{ishihara}\cite{scalapino2}. 
Finally,  experimental data on nanopowders\cite{paturi} also indicate that one should not be overly 
pessimistic about  cluster calculations. YBa$_{2}$Cu$_{3}$O$_{6+x}$ powders 
consisting of  
0.7 nm thick grains with a 11 nm radius show the same T$_{C}$ as the 
bulk,  indicating that going to the thermodynamic limit could not be 
essential to understand pairing.  
Cluster calculations  can be relevant 
and physically 
insightful concerning the interplay of various degrees of freedom on 
pair structure and formation.

}

\end{document}